\documentclass[twocolumn]{revtex4}  

\usepackage[english]{babel}

\usepackage{dcolumn}
\usepackage{subfigure}
\usepackage{multirow}
\usepackage{amsmath,amsfonts,amssymb}

\usepackage{graphicx}  
\usepackage{latexsym}   

\newcommand{\N}{\text{${\rm{I\!N}}$}}

\begin{document}

\title{Electromagnetic energy within magnetic spheres}

\author{\firstname{Tiago}  Jos\'e  \surname{Arruda}}
\affiliation{Faculdade de Filosofia,~Ci\^encias e Letras de Ribeir\~ao Preto,\\
             Universidade de S\~ao Paulo \\
             Avenida Bandeirantes, 3900 \\
             14040-901, Ribeir\~ao Preto, S\~ao Paulo, Brazil}

\author{\firstname{Alexandre} Souto \surname{Martinez}}
\email{asmartinez@ffclrp.usp.br}
\affiliation{Faculdade de Filosofia,~Ci\^encias e Letras de Ribeir\~ao Preto,\\
             Universidade de S\~ao Paulo \\
             Avenida Bandeirantes, 3900 \\
             14040-901, Ribeir\~ao Preto, S\~ao Paulo, Brazil}
\affiliation{National Institute of Science and Technology in Complex Systems}
%
%

\begin{abstract}
Consider that an incident plane wave is scattered by a homogeneous
and isotropic magnetic sphere of finite radius. We determine, by
means of the rigorous Mie theory, an exact expression for the
time-averaged electromagnetic energy within this particle. For
magnetic scatterers, we find that the value of the average
internal energy in the resonance picks is much larger than the one
associated with a scatterer with the same nonmagnetic medium
properties. This result is valid even, and specially, for low size
parameter values. Expressions for the contributions of the
radial and angular field components to the internal energy are
determined. For the analytical study of the weak absorption
regime, we derive an exact expression for the absorption cross
section in terms of the magnetic Mie internal coefficients. We
stress that although the electromagnetic scattering by particles
is a well-documented topic, almost no attention has been devoted
to magnetic scatterers. Our aim is to provide some new analytical
results, which can be used for magnetic particles, and emphasize
the unusual properties of the magnetic scatters, which could be
important in some applications.

\end{abstract}




\maketitle


\section{Introduction}

The research in magneto-optics, both theoretical and experimental,
has been mainly devoted to the study of magnetic properties of
thin films. Magneto-optical effects are characterized by the
change in the state of light polarization in the presence of
magnetic materials, both in transmission (Faraday effect) and
reflection (Kerr effect). Brillouin light scattering technique
allows the investigation of spin waves in magnetic films and
layered structures through the light scattering by magnons. Here,
we are concerned with another feature in the magneto-optics
research: the electromagnetic (EM) scattering by magnetic
particles \cite{kerkermag,alexandre0,alexandre,alexandre1,alexandre2}.
Although the EM scattering by particles is a well-documented
topic~\cite{bohren,hulst,barber,kerker,stratton}, little attention
has been given to the case of EM scattering by magnetic particles.
Recently, it has been a growing interest on photonic band gap
materials (PBGs) made of ferromagnetic materials, like soft
ferrites, at microwave or radio frequencies
\cite{sigalas,ping,liu,lin2}.
Other important applications involving
magnetic materials, such as microwave filters, metamaterials, high
density magnetic recording media, have been reported
\cite{lin,tarento}. here, the approach we follow
is the classical one for single Mie scattering
\cite{bohren,kerker,hulst,stratton}, in which no applied external
field is considered.

The EM radiation scattering by magnetic spheres is described on
the basis of the Mie theory, in which an incident plane wave, with
wavenumber $k$, is scattered by a homogeneous sphere of radius
$a$. We assume that both the scatterer and the medium are
nonmagneto-optical active and that the incident radiation is a
vetorial wave. In general, the bulk of analysis takes place in
far-field approximation, ignoring the evanescence and the internal
fields in the scattering center \cite{bohren,kerker,hulst}. The
interest, therefore, lies in the behavior of the scattered fields
and all the quantities of interest to describe EM scattering by
spherical particles, such as cross sections and the anisotropy
factor $\langle \cos \theta \rangle$ (i.e., the mean value of the
cosine of the scattering angle $\theta$), can be expressed in
terms of the Mie coefficients $a_{n}$ and $b_{n}$ \cite{kerker}.
For magnetic scatterers, in particular, $a_{n}$ and $b_{n}$ have
been obtained by Kerker et al.~\cite{kerkermag}. Nevertheless,
here, as in the original work of Bott and Zdunkowski~\cite{bott}
for nonmagnetic spheres, the internal fields in Mie single
scattering gain special attention and some related quantities are
studied.

Bott and Zdunkowski~\cite{bott} present the exact and approximate
analytical expressions for the time-averaged EM energy within a
dielectric sphere. The calculations have been anchored on the
rigorous Mie theory, and the expressions have been derived, as
usual, with the assumption of equality between the magnetic
permeability tensors of medium and particle. This configuration is
denominated \emph{nonmagnetic scattering} \cite{kerkermag}. It is
pointed out in \cite{bott} that those calculations are of
importance for the study of photochemical reactions within
atmospheric water spheres.

The aim of this paper is to provide a detailed description of the
time-averaged EM energy within magnetic particles (assumed to be
spherical), emphasizing their unusual properties, which in turn
could be explored in  as microwave filters and PBGs~\cite{lin} or
in the search of photon localization in the multiple scattering
regime~\cite{alexandre1,alexandre2}. In Sec.~\ref{sec:theory} is
presented a brief resume about the construction of the exact
solution in the single magnetic Mie scattering and its principal
analytical results. Both the EM internal fields and the magnetic
Mie coefficients are presented. The determination of the exact
expression for the time-averaged EM energy within a magnetic
scatterer is shown in Sec.~\ref{sec:energy}. The problem symmetry
allows us to express separately the contribution of the radial and
angular components to the average internal energy. A new
expression for the absorption cross section in the magnetic case
is determined. To validate our expressions, for instance, we
determine the same particular relations studied in \cite{bott}.
Special attention is paid to our approach concerning the
differences to \cite{bott}. Finally, we present some numerical
results in Sec.~\ref{sec:numerical}. We compare the magnetic and
nonmagnetic scattering. The basic relations involving the Bessel
and associated Legendre functions are presented in the
Appendix~\ref{sec:special}. Those expressions are important to the
calculation of the quantities related to the time-averaged
internal energy. In the Appendix~\ref{sec:limiting}, some
classical limiting cases are considered and we give a set of
approximated magnetic Mie coefficients.

\section{Analytical calculation of scattering quantities}
\label{sec:theory}

To deal with EM wave scattering by a single particle embedded in a
medium, one must assume some special features for the medium and
the incident wave. Among these assumptions, the particle is
considered isolated in an infinite medium, which allows one to
ignore the effect of multiple scattering \cite{bohren,hulst}. Both
particle and medium are considered linear, homogeneous, and
isotropic, having inductive capacities ($\epsilon_1,\mu_1$) and
($\epsilon,\mu$), respectively. Thereby, once we assume the media
are nonmagneto-optic active, those tensors, respective to magnetic
permeability ($\mu$) and electric permittivity ($\epsilon$), can
be expressed by a scalar quantity times an unitary tensor. In
particular, it is assumed that there are absorptive components
within the scatterer, so the quantities $\epsilon_1$ and $\mu_1$
are complex.

The incident radiation is considered plane, monochromatic, and
polarized complex EM wave, which is expressed as
\begin{equation}
\mathbf{E}_i(\mathbf{r},t) = \mathbf{E}_0
\exp\left[\imath\left(\mathbf{k \cdot r} -\omega t\right)\right],
\end{equation}
with wave amplitude $\mathbf{E}_0=E_0\mathbf{e}_x$, wave vector
$\mathbf{k} = k \mathbf{e}_{z}$ and angular frequency
$\omega=2\pi/\lambda$, where $\lambda$ is the wavelength. Due to
the spherical symmetry of the scattering center, there is no loss
of generality taking the electric field polarized on the $x$ axis
direction. Also, the linearity of the macroscopic Maxwell
equations and Fourier theory allow one to generalize this
monochromatic case to a polychromatic one \cite{bohren}.

The incident, scattered and internal vector waves have the same
angular frequency $\omega$, once we are not accounting for
possible energy variations in the interaction with the scatterer.
Thus, quantum fluctuations such as in Raman scattering is
neglected, and a classical description is
adopted~\cite{bohren,hulst}.

In the rigorous Mie theory it is quite common to assume the
equality between the magnetic permeability tensors of the particle
and medium. This consideration ignores the most general case in
which these complex tensors are different. The absolute value of
the magnetic permeability $\mu_1$ can assume values much larger
than $\mu$, as in the case of soft ferromagnetic particles in the
microwave range, for instance \cite{alexandre0,sigalas}. In this
present work, the Mie coefficients are recalculated in this
general case, referred to as \emph{magnetic
scattering}~\cite{kerkermag,alexandre,alexandre1,alexandre2}, and the associated Mie
coefficients of the internal fields, which have not been studied
so far, are presented. The expressions here obtained are valid for
a wide class of soft ferrites with magnetic loss. The assumption
of the isotropic magnetic permeability (and electric permittivity)
allows one to solve the scattering problem in a simple way.
However, to lower the magnetic loss of these magnetic materials,
it is usually to consider them in presence of an applied external
magnetic field \cite{sigalas,ping,lin,lin2,tarento}. In this
situation, the relative magnetic permeability is anisotropic and
its tensor elements are depends on this externally applied field.

From the Maxwell theory, we have that a time-harmonic EM field
($\mathbf{E}$, $\mathbf{H}$) in an homogeneous, isotropic and
linear medium must satisfy the vectorial Helmholtz equation
$[\mathbf{\nabla}^2+k^2]\mathbf{E}=\mathbf{0},\
[\mathbf{\nabla}^2+k^2]\mathbf{H}=\mathbf{0}$, where
$k^2=-\epsilon\mu\partial_t^2=\omega^2\epsilon\mu$, and be
divergence-free null: $\mathbf{\nabla\cdot E}=\mathbf{\nabla\cdot
H}=0$. The quantity $|\mathbf{k}|=k$ is the wavenumber and it is
related to the travelling wave. In addition, $\mathbf{E}$ and
$\mathbf{H}$ are not independent:
$\mathbf{\nabla}\times\mathbf{E}=\imath\omega\mu\mathbf{H}$,
$\mathbf{\nabla}\times\mathbf{H}=-\imath\omega\epsilon\mathbf{E}$.
To simplify the resolution of these equations, we build solutions
which are dependent of a scalar function $\psi$, called a
generating function for the vector harmonics \cite{bohren,hulst}.
In this particular case, once the symmetry of the problem is
spherical, the solutions of the equations above are the spherical
vector harmonics, expressed by
$\mathbf{M}=\nabla\times\left(\mathbf{r}\psi\right)$ and
$\mathbf{N}=\nabla\times\mathbf{M}/k$. The imposition of the
vector harmonics are solutions of the Maxwell equations implies
that $ [\mathbf{\nabla}^2+k^2]\psi={0}$. Thus, the problem of the
scattered waves by a spherical particle resumes to solve this
scalar Helmholtz equation in spherical coordinates.

Another way to tackle this problem without considering the vector
harmonics employs the Hertz potential \cite{kerker,ishimaru}.
However, we prefer to adopt the same framework of Bohren and
Huffman~\cite{bohren}, in which the plane waves are directly
expanded in terms of spherical vector harmonics. The solution of
the scalar Helmholtz equation is
$\psi_{nm}(kr,\cos\theta,\phi)=z_n(kr)P_n^m(\cos\theta)\exp(\imath
   m\phi)$, where $z_n(kr)$ is a generic Bessel spherical function and $P_n^m(\cos \theta)$ are
associated Legendre functions, $n$ natural and $m$ integer. By
means of $\psi_{nm}$, we can readily derive the spherical vector
harmonics above defined. From the expansion of the incident fields
in terms of $\mathbf{M}_{nm}$ and $\mathbf{N}_{nm}$, we find that
only $m=1$ contributes to this new representation due to the
spherical symmetry of the scatterer \cite{bohren,hulst}. Using the
boundary conditions of this problem, we can express the internal
and the scattered fields in terms of spherical vector harmonics.
In special, the coefficients of these expansions are referred to
as the \emph{Mie coefficients}. They provide the information about
the interaction between the incident wave and the spherical
particle. Explicitly, the boundary condition ($r = a$) is
expressed by
$(\mathbf{E}_i+\mathbf{E}_s-\mathbf{E}_1)\times\mathbf{e}_r=(\mathbf{H}_i+\mathbf{H}_s-\mathbf{H}_1)\times\mathbf{e}_r=\mathbf{0}$,
where 1 is the index related to the particle (internal fields),
and $i$ and $s$ refer to the incident and scattered fields,
respectively; $\mathbf{e}_r$ is the radial unity vector in the
polar spherical coordinate system.

\subsection{Internal fields}
\label{sec:fields}

Assuming the incident EM wave is polarized in the $\mathbf{e}_x$
direction, and the scattering center is placed at the origin of
the coordinate system, we obtain an expression for the expansion
of this field in terms of spherical harmonics. Imposing on the
boundary between the sphere and the surrounding medium the
continuity of the EM fields $-$ in fact, their (electrical and
magnetic) tangential components $-$, expressions are determined
for the internal and scattered fields~\cite{bohren,hulst,kerker}.

Using the same notation of Bohren and Huffman~\cite{bohren}, we
can give the components of the electric and magnetic vectors,
$\mathbf{E}_1$ and $\mathbf{H}_1$, respectively, of the interior
field in a spherical coordinate system $(r,\theta,\phi)$ by
\begin{eqnarray*}
        E_{1r}&=&\displaystyle \frac{- \imath \cos\phi\sin\theta}{\rho_1^2}\sum_{n=1}^{\infty}E_nd_n\psi_n(\rho_1)n(n+1)\pi_n\
        ,\\
        E_{1\theta}&=&\displaystyle\frac{\cos\phi}{\rho_1}\sum_{n=1}^{\infty}E_n\left[c_n\pi_n\psi_n(\rho_1)-\imath
        d_n\tau_n\psi_n'(\rho_1)\right]\ ,\\
        E_{1\phi}&=&\displaystyle\frac{\sin\phi}{\rho_1}\sum_{n=1}^{\infty}E_n\left[\imath
        d_n\pi_n\psi_n'(\rho_1)-c_n\tau_n\psi_n(\rho_1)\right]\
        ;\\
        H_{1r}&=&\displaystyle\frac{- \imath k_1}{\omega\mu_1}\frac{\sin\phi\sin\theta}{\rho_1^2}\sum_{n=1}^{\infty}E_nc_n\psi_n(\rho_1)n(n+1)\pi_n\
        ,\\
        H_{1\theta}&=&\displaystyle\frac{k_1}{\omega\mu_1}\frac{\sin\phi}{\rho_1}\sum_{n=1}^{\infty}E_n\left[d_n\pi_n\psi_n(\rho_1)-\imath
        c_n\tau_n\psi_n'(\rho_1)\right]\ ,\\
        H_{1\phi}&=&\displaystyle\frac{k_1}{\omega\mu_1}\frac{\cos\phi}{\rho_1}\sum_{n=1}^{\infty}E_n\left[d_n\tau_n\psi_n(\rho_1)-\imath
        c_n\pi_n\psi_n'(\rho_1)\right]\ .
\end{eqnarray*}
with $\rho_1=k_1r$, $E_n=\imath^nE_0({2n+1})/[{n(n+1)}]$, $\pi_n$
and $\tau_n$ are angular functions defined in the appendix
\ref{sec:angular} and $\psi_n(\rho_1)=\rho_1j_n(\rho_1)$ is a
Ricatti-Bessel function. The functions $c_n$ and $d_n$ are the
internal magnetic Mie coefficients, which are presented in the
section below. We outline that, because of the notation (units in
the SI and time factor) here adopted, the internal EM field
($\mathbf{E}_1,\mathbf{H}_1$) is not the same presented in
\cite{bott}.

\subsection{Magnetic Mie coefficients}
\label{sec:coefficients}

If we do not assume the equality between $\mu$ and $\mu_1$ on the
problem boundary condition, we can determine the \emph{magnetic
Mie coefficients} for the scattering ($a_n$ and $b_n$, obtained by
Kerker at al.) and internal ($c_n$ and $d_n$)
fields~\cite{kerkermag,alexandre0,alexandre,alexandre1,alexandre2}.
Explicitly,
\begin{eqnarray}
    a_n&=&\displaystyle\frac{\widetilde{m}\psi_n(mx)\psi_n'(x)-\psi_n(x)\psi_n'(mx)}{\widetilde{m}\psi_n(mx)\xi_n'(x)-\xi_n(x)\psi_n'(mx)}\ ,\label{coef1}\\
    b_n&=&\displaystyle\frac{\psi_n(mx)\psi_n'(x)-\widetilde{m}\psi_n(x)\psi_n'(mx)}{\psi_n(mx)\xi_n'(x)-\widetilde{m}\xi_n(x)\psi_n'(mx)}\ ,\label{coef2}\\
    c_n&=&\displaystyle\frac{m\imath}{\psi_n(mx)\xi_n'(x)-\widetilde{m}\xi_n(x)\psi_n'(mx)}\ ,\label{coef3}\\
    d_n&=&\displaystyle\frac{m\imath}{\widetilde{m}\psi_n(mx)\xi_n'(x)-\xi_n(x)\psi_n'(mx)}\  ,\label{coef4}
\end{eqnarray}
with the assumption that the function domains are restricted in
such a manner that the denominators do not vanish. The quantity
$x=ka$ is the size parameter of the spherical particle, being $a$
its radius and $k=|\mathbf{k}|$ the wavenumber of incident and
scattered waves, and $\xi_n(x)=x[j_n(x)+\imath y_n(x)]$ is the
Ricatti-Hankel function of first kind. In addition,
$m=({\mu_1\epsilon_1/\mu\epsilon})^{1/2}$ is the relative
refraction index and
$\widetilde{m}=({\mu\epsilon_1/\mu_1\epsilon})^{1/2}$ is the
relative impedance between the media. For $\mu=\mu_1$, then
$\widetilde{m}=m$ and the usual expressions for the Mie
coefficients (\ref{coef1}-\ref{coef4}) are
recovered~\cite{bohren,kerker,hulst}.

There are some notation differences between this work and the one
presented by Bott and Zdunkowski~\cite{bott}. Here, we use the
same framework of Bohren and Huffman~\cite{bohren}, which have
treated the scattering problem of light by means of International
System of Units (SI), and have adopted $\exp({-\imath\omega t})$
as time-harmonic dependency for the fields. Otherwise, \cite{bott}
have used the same notation as van de Hulst~\cite{hulst}, which
has dealt with the scattering problem in the Gaussian System of
Units, and has adopted $\exp({\imath\omega t})$ as time-harmonic
dependency. These approach differences appear explicitly in the
choice of the Hankel functions, which is strictly associated with
the asymptotic limit for the scattered fields (the well-known
far-field approximation), and consequently determines the
dependencies of the Mie coefficients. Another difference between
these representations is related to the signal of the imaginary
part of the relative refraction index $m=m_r+\imath m_i$, which is
positive in the framework we have chosen \cite{bohren,hulst}.

\section{Time-averaged electromagnetic energy}
\label{sec:energy}

For a linear, homogeneous, and isotropic medium, the classical
theory of electromagnetism provides an expression for the
time-averaged EM energy as an integral of the component
intensities within the volume under analysis. In the case of a
spherical particle with radius $a$ and internal complex EM field
$(\mathbf{E}_1,\mathbf{H}_1)$, we have the following
relation~\cite{bott,griffiths}:
\begin{eqnarray}
    W(a)&=& \int_0^{2\pi} {\rm d}\phi \int_{-1}^{1} {\rm
    d}(\cos\theta) \int_0^a {\rm d}r r^2\nonumber\\
    & &\ \times{\rm
    Re}\bigg[\frac{\epsilon_1}{4}\left(|E_{1r}|^2+|E_{1\theta}|^2+|E_{1\phi}|^2\right)\nonumber\\
    & &\quad+\frac{\mu_1}{4}\left(|H_{1r}|^2+|H_{1\theta}|^2+|H_{1\phi}|^2\right)\bigg]\ .\label{ener}
\end{eqnarray}

In Eq.~(\ref{ener}), with respect to the field representations, we
outline that the permutation between a definite integral and a sum
of an infinite series is not trivial. In the following
calculations, we are not concerned about showing explicitly each
one of the simplifications; we just assume that the function
series related to the field intensities converge uniformly in the
domain $0< r\leq a$, $0\leq\theta\leq\pi$, $0<\phi\leq2\pi$.
Obviously, this mathematical condition is in agreement that the
energy within a finite sphere is also finite.

\subsection{Electric and magnetic internal fields}

Looking closely to the definition of Eq.~(\ref{ener}), one can ask
about the contribution to the total internal energy associated with
electric and magnetic fields separately, or even about the
contribution of their fields components in spherical coordinates
$(r,\theta,\phi)$ to this average energy. These questions have not
addressed by Bott and Zdunkowski in their \cite{bott} paper.

\subsubsection{Radial component}

From Eq.~(\ref{ener}), we obtain that the contribution of the
radial component associated with the electric field to the internal
energy is given by
\begin{equation}\begin{split}
    W_{rE}(a)&=\frac{{\rm Re}(\epsilon_1)}{4}\int_{-1}^{1} {\rm
    d}(\cos\theta) \int_0^{2\pi} {\rm d}\phi \int_0^a {\rm d}r r^2 \big|E_{1r}\big|^2\\
    &=\frac{\pi}{4}{\rm Re}(\epsilon_1)\int_0^a {\rm d}r \sum_{n=1}^{\infty}\sum_{n'=1}^{\infty}E_nE_{n'}^*j_n(\rho_1)j_{n'}^*(\rho_1)\frac{d_nd_{n'}^*}{|k_1|^2}\\
    &\quad\quad\times nn'(n+1)(n'+1)\underbrace{\int_{-1}^{1} {\rm
    d}(\cos\theta) \sin^2\theta\pi_n\pi_{n'}}_{\rm Eq.~(\ref{ort3})}\\
    &=\frac{\pi}{2} |E_0|^2\frac{{\rm Re}(\epsilon_1)}{|k_1|^2}\sum_{n=1}^{\infty}n(n+1)(2n+1)\\
    &\qquad\qquad\qquad\qquad\qquad\times|d_n|^2\int_0^a {\rm
    d}r |j_n(\rho_1)|^2\ .\label{w-er}
\end{split}\end{equation}

Proceeding in the same way, one derives an analogous expression
for the radial component related to the magnetic field:
\begin{eqnarray}
    W_{rH}(a)&=&\frac{\pi}{2} |E_0|^2\frac{{\rm Re}(\mu_1^{-1})}{\omega^2}\sum_{n=1}^{\infty}n(n+1)\nonumber\\
    & &\times(2n+1)|c_n|^2\int_0^a {\rm
    d}r |j_n(\rho_1)|^2\ .\label{w-hr}
\end{eqnarray}

An important point to be noted here is that the integral above
cannot be simplified by means of Eq.~(\ref{int-bes1}) and
recurrence relations presented in Appendix \ref{sec:radial}.

\subsubsection{Angular components}

Because of spherical symmetry of the system, it is not possible to
write the contributions of the angular and azimuthal components to
internal energy separately. If one tries to do that, the necessary
relations to simply the \emph{double sums}, as exemplified in
Eq.~(\ref{w-er}), do not appear. Fortunately, if one considers
both ($\theta,\phi$) contributions to internal energy, these
relations are not lost. Thus, using the relations (\ref{ort1}) and
(\ref{ort2}) from Appendix~\ref{sec:angular} and the first term of
Eq.~(\ref{ener}), it follows that the time-averaged energy
associated with angular components of the electric field is
expressed by
\begin{equation}\begin{split}
    [W_{\theta E}+W_{\phi E}](a)& = \frac{{\rm Re}(\epsilon_1)}{4} \int_{-1}^{1} {\rm d} (\cos \theta) \int_0^{2\pi} {\rm d}\phi \\
                                &\quad\times\int_0^a {\rm d}r r^2  \left(|E_{1\theta}|^2 +|E_{1\phi}|^2\right)\\
    &= \frac{\pi}{2}|E_0|^2\frac{{\rm
    Re}(\epsilon_1)}{|k_1|^2}\sum_{n=1}^{\infty}(2n+1)\\
    &\quad\times\int_0^a {\rm d}r \left(|c_n\psi_n(\rho_1)|^2+|d_n\psi_n'(\rho_1)|^2\right).
    \label{w-ea}
\end{split}\end{equation}

Similarly, for the magnetic field we obtain
\begin{equation}\begin{split}
    [W_{\theta H}+W_{\phi H}](a)&=\frac{\pi}{2}|E_0|^2\frac{{\rm
    Re}(\mu_1^{-1})}{\omega^2}\sum_{n=1}^{\infty}(2n+1)\\
    &\quad\times\int_0^a {\rm
    d}r \left(|d_n\psi_n(\rho_1)|^2+|c_n\psi_n'(\rho_1)|^2\right).\label{w-ha}
\end{split}\end{equation}

Here, the same problem of the expression $W_r(a)$ arises. Whereas
the integral of Ricatti-Bessel function is only another way to
write Eq.~(\ref{int-bes1}), the second integral above cannot be
simplified.

\subsection{Time-averaged internal energy}

From the expressions obtained in the previous section for each one
of the internal field components, we can calculate the total
time-averaged energy inside the sphere. For the internal electric
field, the expression is
\begin{equation}\begin{split}
    W_{E}(a)&=W_{rE}(a)+[W_{\theta E}+W_{\phi E}](a)\\
    &=\frac{3}{4}W_0{\rm
    Re}(m\widetilde{m})\sum_{n=1}^{\infty}\big\{(2n+1) |c_n|^2 \mathcal{I}_n(y) \\
    &\quad+ |d_n|^2 \big[n \mathcal{I}_{n+1}(y)+(n+1)\mathcal{I}_{n-1}(y) \big]\big\} \;
    ,\label{w-e}
\end{split}\end{equation}
where
\begin{equation}
\mathcal{I}_n(y)= \frac{1}{a^{3}} \; \int_0^a {\rm d}r r^2\big|j_n(\rho_1)\big|^2
\end{equation}
is given by Eq.~(\ref{int-bes1}) and $W_0$ denotes the
time-averaged EM energy of a sphere with radius $a$ having the
same EM properties of the surrounding medium:
\begin{eqnarray}
    W_0     =  \frac{2}{3}\pi a^3|E_0|^2\epsilon\ .\label{w-0}
\end{eqnarray}
For sake of simplicity, the dependence of $\mathcal{I}_n$ with
respect to $y^*=m^*ka$, like the case of function $W=W(a,y,y^*)$,
is omitted.

In the same way, for the internal magnetic field, the average
internal energy is given by
\begin{equation}\begin{split}
    W_{H}(a)&=W_{rH}(a)+[W_{\theta H}+W_{\phi H}](a)\\
    &=\frac{3}{4}W_0{\rm
    Re}(m\widetilde{m}^*) \sum_{n=1}^{\infty}\big\{(2n+1) |d_n|^2 \mathcal{I}_n(y)\\
    &\quad+|c_n|^2\big[n\mathcal{I}_{n+1}(y)+(n+1)\mathcal{I}_{n-1}(y) \big]\big\} \;
    .\label{w-h}
\end{split}\end{equation}

Once we have expressions for electric and magnetic energy within a
sphere, it is possible to determine the expression for the total
time-averaged EM energy inside the scatterer:
$W(a)=W_{E}(a)+W_{H}(a)$. Explicitly,
\begin{equation}\begin{split}
    W(a)&=\frac{3}{4}W_0 \sum_{n=1}^{\infty}|\psi_n(y)|^{-2}\big[n\beta_n\mathcal{I}_{n+1}(y)\\
    &\quad +(n+1)\beta_n\mathcal{I}_{n-1}(y)+(2n+1)\alpha_n\mathcal{I}_n(y)\big]\
    ,\label{ener-first}
\end{split}\end{equation}
where
\begin{eqnarray}
    \alpha_n &=& {|\psi_n(y)|^2}\left[{\rm Re}(m\widetilde{m})|c_n|^2+{\rm
    Re}\left(m\widetilde{m}^*\right)|d_n|^2\right],\label{alphan} \\
    \beta_n &=& {|\psi_n(y)|^2}\left[{\rm Re}(m\widetilde{m})|d_n|^2+{\rm
    Re}\left(m\widetilde{m}^*\right)|c_n|^2\right] .\label{betan}
\end{eqnarray}

Also, to obtain analogous expressions to the ones presented in
\cite{bott}, Eq.~(\ref{ener-first}) can be rewritten as
\begin{eqnarray}
    W(a) &=&\frac{3}{4}W_0\sum_{n=1}^{\infty}\frac{2n+1}{y^2-y^{*2}}\bigg\{\alpha_n\left[\frac{A_n(y^*)}{y}-\frac{A_n(y)}{y^*}\right]\nonumber\\
    & &+\beta_n\left[\frac{A_n(y^*)}{y^*}-\frac{A_n(y)}{y}\right]\bigg\}\
    ,\label{ener-med}
\end{eqnarray}
with $y=mka$ and $A_n(y) = d_y \ln \psi(y)$.

\subsection{Dielectric sphere}
\label{sec:dielectric}

A particular situation to be considered here refers to a
dielectric sphere, which has studied by Bott and Zdunkowski
\cite{bott}. With this aim, consider the nonmagnetic case, i.e.,
$\mu=\mu_1$. Thereby, it results that
$m=\widetilde{m}=({\epsilon_1/\epsilon})^{1/2}$. With this
assumption, note that ${\rm Re}\left(m^2\right)
=({m^2+m^{*2}})/{2}$ and ${\rm Re}(mm^*) = |m|^2$. Substituting
these into Eqs.~(\ref{alphan}) and (\ref{betan}), the expression
for the internal energy obtained by Bott and Zdunkowski is
recovered. Once again, we emphasize that our notation is not the
same that is employed in \cite{bott}. Indeed, we can recover the
same results by means of the following substitutions: $\xi_n(x)
\dashrightarrow  \zeta_n(x)$, $c_n \dashrightarrow  m d_n$, $d_n
\dashrightarrow m c_n$, and assuming $m=\widetilde{m}$. Here,
$\zeta_n(x)=x[j_n(x)-\imath y_n(x)]$ is the Ricatti-Hankel
function of second kind, which is related to the choice of the
time-harmonic dependence for the EM fields, like it is mentioned
in the beginning of this
description~\cite{bohren,hulst,kerker,barber,stratton}.

Employing the recurrence relations involving Bessel spherical
functions \cite{watson,arfken}, we obtain the derivative of first
order $A_n'(y)=-1-A_n^2(y)+{n(n+1)}/{y^2}$. Therefore, using the
L'Hospital rule, the limiting case of a perfect dielectric sphere,
which takes place when $m_i\to0$, provides $ 4 y^2
\lim_{m_i\to0}W(a)= 3 W_0\sum_{n=1}^{\infty} \gamma_n (2n+1)
[1+A_n^2(y)-n(n+1)/y^2]$, where
$\gamma_n=m^2|\psi_n(y)|^2(|c_n|^2+|d_n|^2)$. Unless some
commented notation differences, this result is the same obtained
in \cite{bott}.

\subsection{Absorption cross section}
\label{sec:absorption}

The classical Mie theory provides a set of useful expressions to
calculate the scattering, total and absorption cross-sections in
the scattering process. Explicitly, call $\sigma_{\rm sca}$ the
scattering cross section and $\sigma_{\rm tot}$ the extinction (or
total) cross section. Using the same framework of \cite{bohren},
one can write
\begin{eqnarray}
\sigma_{\rm
sca}&=&\frac{2\pi}{k^2}\sum_{n=1}^{\infty}(2n+1)\left(|a_n|^2+|b_n|^2\right)\
,\label{sca}\\
\sigma_{\rm tot}&=&\frac{2\pi}{k^2}\sum_{n=1}^{\infty}(2n+1){\rm
Re}\left\{a_n+b_n\right\}\ .\label{tot}
\end{eqnarray}

Consequently, the absorption cross section $\sigma_{\rm abs}$
associated with the scatterer is defined in terms of $\sigma_{\rm
sca}$ and $\sigma_{\rm tot}$ by the relation: $\sigma_{\rm abs}=\sigma_{\rm tot}-\sigma_{sca}$.
In other words, the absorption cross section in the Mie single
scattering is determined by quantities and coefficients related
only to the scattered EM fields \cite{bohren,hulst,kerker}.
Although it is suitable and even natural to express the absorption
cross section in terms of the internal coefficients $c_n$ and
$d_n$, notice that we do not do any reference to the internal EM
fields.

From the boundary conditions in the sphere problem \cite{bohren},
the Mie coefficients are linked by the equations below:
\begin{eqnarray}
    h_n^{(1)}(x)b_n&=&j_n(x)-j_n(mx)c_n\ ,\\
    h_n^{(1)}(x)a_n&=&j_n(x)-\widetilde{m}j_n(mx)d_n\ .
\end{eqnarray}
Thus, substituting the coefficients $a_n$ and $b_n$ into
$\sigma_{\rm abs}=\sigma_{\rm tot}-\sigma_{sca}$ and manipulating that, we obtain the following expression respective to absorption cross section:
\begin{eqnarray}
    \sigma_{\rm abs}&=&\frac{2\pi}{k^2}\sum_{n=1}^{\infty}(2n+1)\bigg\{{\rm Re}\left[\frac{\psi_n(mx)}{m\xi_n^*(x)}(c_n+\widetilde{m}d_n)\right]\nonumber\\
    & &-\frac{\left|\psi_n(mx)\right|^2}{\left|m\xi_n(x)\right|^2}\left(|c_n|^2+\left|\widetilde{m}d_n\right|^2\right)\bigg\}\
    .
\end{eqnarray}
Finally, using the definition of $c_n$ and $d_n$ given by
Eqs.~(\ref{coef3}) and (\ref{coef4}) and the fact that ${\rm
    Re}\left[-i\xi_n^*(x)\xi_n'(x)\right]=\chi_n(x)\psi_n'(x)-\psi_n(x)\chi_n'(x)=1$, where $\chi_n(x)=-xy_n(x)$ is the Ricatti-Neumann function, we obtain
\begin{eqnarray}
    \sigma_{\rm abs}&=&\frac{2\pi}{k^2}\sum_{n=1}^{\infty}(2n+1)\left(|c_n|^2+|d_n|^2\right)\nonumber\\
    & &\qquad\quad\quad\quad\times{\rm
Im}\left[\frac{\widetilde{m}}{|m|^2}\psi_n(y)\psi_n'(y^*)\right]\
.\label{sigma-abs}
\end{eqnarray}
The exact expression (\ref{sigma-abs}) for the magnetic absorption
cross section in terms of the internal Mie coefficients is not
found in the classical books of the scattering theory~\cite{bohren,hulst,kerker} and it had not been determined so far.

\section{Numerical results}
\label{sec:numerical}

In this section we present some numerical analysis from the exact
equations determined in the sections above. Our aim is not to
restrict our studies in some particular case of magnetic
scattering, but to introduce a general formulation of the internal
energy which can be used whether in magnetic case
($\mu\not=\mu_1$) or in nonmagnetic one ($\mu=\mu_1$). Here, all
numerical results are obtained by means of a program created by us
using the free software for scientific computation $Scilab\
5.1.1$. As usual in the numerical Mie scattering, rather than to
do infinite sums in $\sum_{n=1}^{\infty}$ in the calculation of
the scattering quantities, which is impossible, we assume an
approximation: finite sums with upper limit $n_{\rm
max}=x+4x^{1/3}+2$, where $x$ is the size parameter~\cite{barber}.

\subsection{Normalized quantities}

For numerical studies, it is suitable to define dimensionless
quantities related to the internal energy and the Mie
coefficients:
\begin{eqnarray}
    W_E^{\rm
    norm}(m,\widetilde{m},ka)=\frac{W_E(m,\widetilde{m},ka;\epsilon,a)}{W_0(\epsilon,a)}\ ,\\
    W_H^{\rm
    norm}(m,\widetilde{m},ka)=\frac{W_H(m,\widetilde{m},ka;\epsilon,a)}{W_0(\epsilon,a)}\ ,
\end{eqnarray}
where $W_E$, $W_H$ and $W_0$ are expressed by Eqs.~(\ref{w-e}),
(\ref{w-h}) and (\ref{w-0}), respectively. The dependence of $W_E$
and $W_H$ on the quantity $m^*$ is omitted.

Therefore, one can define the normalization of the total
time-averaged internal energy by the relation $W_{\rm tot}^{\rm
norm}=W_E^{\rm norm}+W_H^{\rm norm}$ or directly from
Eq.~(\ref{ener-med}): $W_{\rm tot}^{\rm
norm}(m,\widetilde{m},ka)=W(a)/W_0$.

Also, from \cite{arfken}, we can obtain the recurrence relation
$(2n+1)^2|j_n(\rho_1)|^2=|\rho_1|^2
\{|j_{n-1}(\rho_1)|^2+|j_{n+1}(\rho_1)|^2+2{\rm
Re}[j_{n-1}(\rho_1)j_{n+1}(\rho_1^*)]\}$;  thus, Eq.~(\ref{w-er})
can be rewritten as
\begin{equation}\begin{split}
    \frac{W_{rE}(a)}{W_0}=&\frac{3}{4}{{\rm Re}(m\widetilde{m})}\sum_{n=1}^{\infty}\frac{n(n+1)}{2n+1}|d_n|^2\\
    &\times\bigg\{\mathcal{I}_{n-1}(y)+\mathcal{I}_{n+1}(y)\\
    &+\frac{2}{a^3}\int_0^a {\rm
    d}r r^2 {\rm Re}\left[j_{n-1}(\rho_1)j_{n+1}(\rho_1^*)\right]\bigg\}\ .
\end{split}\end{equation}
Note that the integral that appears in the sum above is quite
similar to that one expressed in Eq.~(\ref{int-bes1}). Although
this one cannot be simplified by means of Eq.~(\ref{int-bes1}), it
is possible to show numerically that the result of this integral
is proportional to $a^3$. It means that we can study the radial
contribution to the internal energy normalized by $W_0$ using only
the dimensionless parameters $m$, $\widetilde{m}$ and $ka$. The
same argument can be applied to both $[W_{\theta E}+W_{\phi
E}](a)/W_0$ and the analogous expressions respective to the
internal magnetic field, given by Eqs.~(\ref{w-hr}) and
(\ref{w-ha}).

In the situations considered here, the values of ${\rm
Re}(m\widetilde{m})$ and ${\rm Re}(m\widetilde{m}^*)$ are very
close in such a way that $W_E(a)\approx W_H(a)$. Thus, we only
consider the total time-averaged internal energy in our analysis.
Further, we remark that although in these studies the relative
magnetic permeability is assumed to be real, there is no such
restriction in the calculated expressions. For soft microwave
ferrites, a more realistic study should consider the magnetic
loss.

\begin{figure}[tb]
\centering
\includegraphics[angle=0, width=1.0\linewidth]{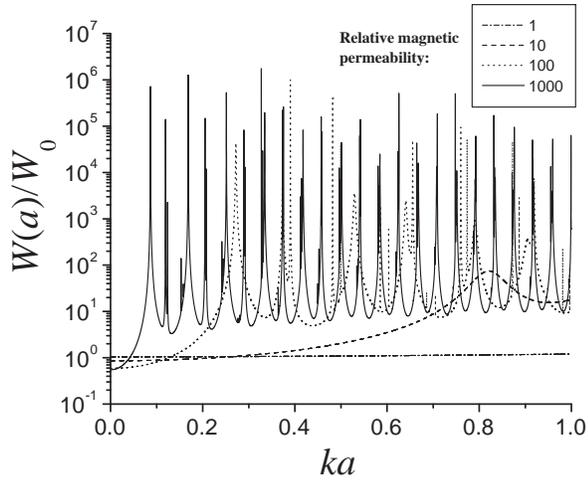}
\caption{Comparison among the distributions of $W(a)/W_0$ in function of the size parameter $ka$. The values of the relative permittivity and permeability related to a non-absorptive sphere are $\epsilon_1/\epsilon=1.4161$ and $\mu_1/\mu=1,10,100,1000$, respectively. The internal energy $W(a)/W_0$ is calculated in the interval $10^{-6}\leq ka\leq1$, $\delta(ka)=10^{-4}$.}
\label{fig1}
\end{figure}

\begin{figure}[tb]
\centering
\includegraphics[angle=0, width=1.0\linewidth]{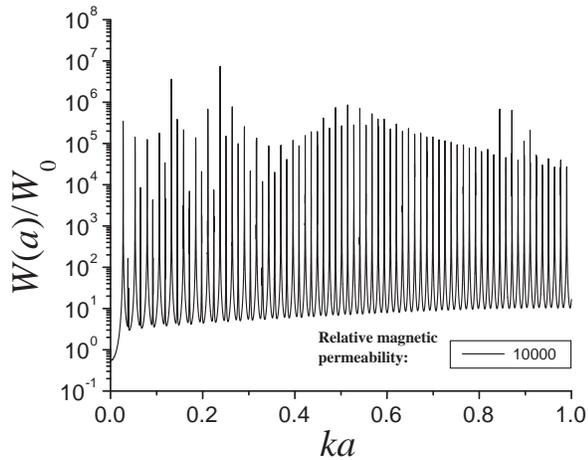}
\caption{The normalized time-averaged internal energy $W(a)/W_0$ within a non-absorptive magnetic sphere plotted as a function of the size parameter $ka$. The values of the relative permittivity and permeability are $\epsilon_1/\epsilon=1.4161$ and $\mu_1/\mu=10^{4}$, respectively. The internal energy $W(a)/W_0$ is calculated in the interval $10^{-6}\leq ka\leq1$, $\delta(ka)=10^{-4}$.}
\label{fig2}
\end{figure}

Figs.~\ref{fig1} and \ref{fig2} illustrate a succession of
narrower picks of the values of the normalized internal energy in
single magnetic Mie scattering as a function of the size parameter
and relative magnetic permeability. Here, we are not concerned
about to study in details the resonances picks and ripple
structure due to the internal coefficients $c_n$ and $d_n$
\cite{bott,petr,dipakbin}. The internal energy of a magnetic
sphere presents resonances peaks even in the limit of small
geometric size (compared to the wavelength). These resonances are
due to the increase of the total cross section due to magnetism.
This increase leads to a decrease of the photon mean free path in
multiple scattering regime in a disordered system. The smaller
mean free path favors the localization phenomenon as pointed in
\cite{alexandre,alexandre1,alexandre2}. For dielectric spheres,
these narrower resonance picks are well known and they are
referred to as morphology dependent-resonances (MDRs)
\cite{michenko}. In the Mie theory, for large size parameters,
these MDRs are commonly observed at the scattered and internal
intensities and at the total cross section.

\begin{figure}[htb]
\centering
\includegraphics[angle=0, width=1\linewidth]{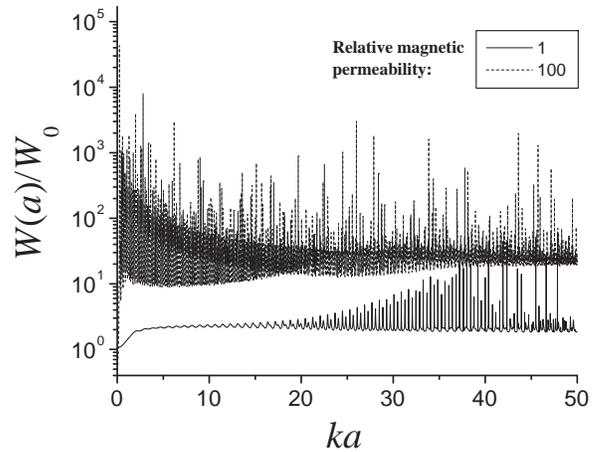}
\caption{The normalized time-averaged internal energy $W(a)/W_0$ plotted as a function of the size parameter $ka$. The values of the relative permeability are $\mu_1/\mu=1$ (nonmagnetic sphere) and $\mu_1/\mu=100$ (magnetic sphere). The relative refraction index is $m=1.334+1.5\times10^{-9}\imath$, which have been used in \cite{bott} in the nonmagnetic scattering approach. The quantities are calculated in the interval $1\leq ka\leq50$, $\delta({ka})=0.01$.}
\label{fig3}
\end{figure}

In our system, observe that when one increases the contribution of
the magnetism in the scatterer, the values of the internal energy
$W(a)$ become much larger than $W_0$, and the narrower picks
appear even for small size parameters. In the nonmagnetic case
reported in \cite{bott}, there is an oppositive tendency, that is,
both the internal energy and the absorption efficiency enhance
with the size parameter. These difference between a nonmagnetic
case and a magnetic one is illustrated in Fig.~\ref{fig3}. This is
due to the increase of the total cross section even though
geometrically the scatterer is much smaller compared to the
wavelength. In other words, the incident EM wave interacts
strongly with the optical cross section instead of the geometrical
one \cite{alexandre0,alexandre,alexandre1,alexandre2}.

\begin{figure}[htb]
\centering
\includegraphics[angle=0, width=1\linewidth]{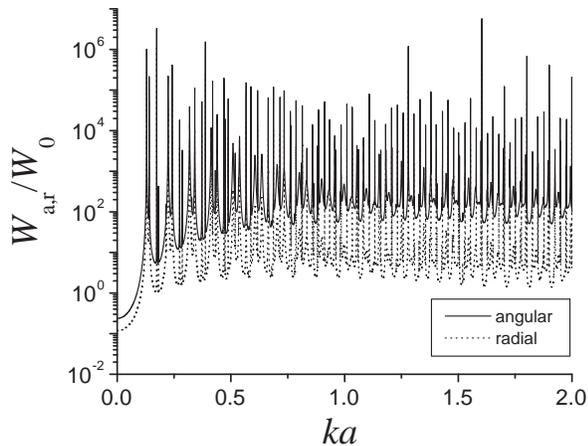}
\caption{The separation of the total time-averaged internal energy in radial and angular contributions respective to both electric and magnetic fields. The values of the relative electric permittivity and magnetic permeability are $\epsilon_1/\epsilon=10$ and $\mu_1/\mu=100$, respectively. The component contributions $W_{r}(a)/W_0$ and $W_{\theta,\phi}(a)/W_0$ are calculated in the interval $10^{-6}\leq ka\leq 2$, $\delta(ka)=10^{-4}$.}
\label{fig4}
\end{figure}

It must be mentioned that this assumption of a non-absorptive
magnetic diffusor is conditioned to the frequency range of the
incident beam. Usually, it is controlled with an external static
magnetic field \cite{sigalas,ping,lin,lin2}. Indeed, there is a
wide variety of soft ferrites which exhibits very large values of
relative magnetic permeability at applied frequencies typically
below 100 MHz with low magnetic loss \cite{sigalas}. For  sake of
simplicity and generality, the situations considered here do not
take into account the scatterer in the magnetized state, and a
scalar value for the $\mu_1/\mu$ is adopted. The dependence on the
angular frequency $\omega$ of the incident EM wave remains
implicit on the value of the size parameter $ka$. Given a value of
$\omega$ for the incident EM wave, a surrounding medium
$(\epsilon,\mu)$ and a scatter $(\epsilon_1,\mu_1)$, one readily
obtains $k=\omega(\mu\epsilon)^{1/2}$ and $k_1=mk$ (see Fig.~\ref{fig4}).

\subsection{Weak absorption regime}
\label{sec:weak}

\begin{figure}[htb]
\centering
\includegraphics[angle=0, width=1\linewidth]{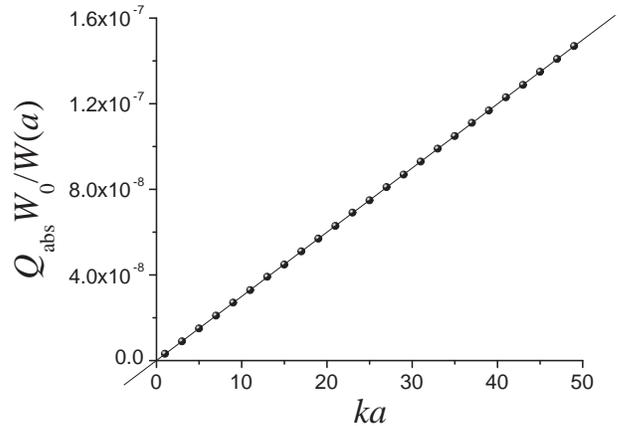}
\caption{Ratio between the absorption efficiency $Q_{\rm abs}$ and the normalized time-averaged internal energy $W(a)/W_0$ plotted as a function of the size parameter $ka$. The values of the relative permeability and refraction index are $\mu_1/\mu=1$ (nonmagnetic sphere) and $m=1.334+1.5\times10^{-9}\imath$, respectively. The quantities are calculated in the interval $1\leq ka\leq 49$, $\delta(ka)=2$. The angular coefficient of linear regression is approximately $2.997\times10^{-9}$, which is in agreement with Eq.~(\ref{WQ}): $8m_i/(3m_r)\approx 2.998\times10^{-9}$.}
\label{fig5}
\end{figure}

In the weak absorption regime (wa), it is quite evident the
relation of the time-averaged internal energy and the absorption
efficiency when we compare them, as it is shown in the
Fig.~\ref{fig5}. This correlation between these quantities in
nonmagnetic scattering have been studied in \cite{bott}.

Analytically, for $m_i\ll m_r$ and $\widetilde{m}_i\ll
\widetilde{m}_r$, we can write $y^2-y^{*2}\approx4\imath
x^2m_rm_i$, ${\rm Re}(m\widetilde{m})\approx m_r\widetilde{m}_r$
and ${\rm Re}\left(m\widetilde{m}^*\right)\approx
m_r\widetilde{m}_r$. Using these approximations in
Eq.~(\ref{ener-med}), it follows that
\begin{eqnarray}
    W_{\rm wa}(a)&\approx&\frac{3}{8}W_0\frac{m_r}{m_i}\frac{2\widetilde{m}_r}{x^3m_r^2}\sum_{n=1}^{\infty}(2n+1)\nonumber\\
    & &\times\left(|c_n|^2+|d_n|^2\right){\rm
    Im}\left[\psi_n(y)\psi_n'(y^*)\right]\ .
\end{eqnarray}
Once the absorption efficiency in the Mie single scattering is
defined by $Q_{\rm abs}=\sigma_{\rm abs}/\sigma_{\rm g}$, where
$\sigma_{\rm g}=\pi a^2$ is the geometric cross section and
$\sigma_{\rm abs}$ is expressed in Eq.~(\ref{sigma-abs}), we can
write that
\begin{eqnarray}
    W_{\rm wa}(a)\approx\frac{3}{8}W_0\frac{m_r}{m_ix}Q_{\rm abs}\
    ,\label{WQ}
\end{eqnarray}
which is the same relation obtained in \cite{bott} in the
nonmagnetic case. Indeed, this approximation is valid wherever
$m_i\ll m_r$ and $\widetilde{m}_i\ll\widetilde{m}_r$; it is not
affected by the value of $\mu_1/\mu$.

In addition, for the case in which $m_r\approx1$, one obtains
$W_{\rm wa}(a)\approx W_0$. Therefore, in this particular
situation, one can write $Q_{\rm abs}\approx 8x/(3m_i)$, which is
a well-known expression \cite{bott,hulst}.

\section{Conclusion}

In this paper we generalize the exact expression of the
time-averaged EM internal energy, obtained firstly in \cite{bott},
to the case of magnetic spherical scatterers. Using the same
framework of \cite{bohren} and assuming the magnetic scattering
approach \cite{kerkermag}, we determine analytical expressions for
the contributions to the EM internal energy related to the fields
components separately. The expressions for the EM internal energy
within a dielectric sphere and the relation derived in \cite{bott}
in the weak absorption regime between the internal energy and the
absorption efficiency are recovered. In special, we find that the
magnetism of the particle does not break the linear relation
between the absorption efficiency and size parameter. To do so, we
analytically calculate an expression for the absorption
efficiency, which depends only on the internal magnetic Mie
coefficients. In addition, we calculate the limiting cases of the
magnetic Mie coefficients and present some important properties of
the radial functions which are used to simply the obtained
expressions. Finally, the main result of this work is that, even
for small scatterers compared to the wavelength, the value of the
EM internal energy within a magnetic sphere is much larger than
that one associated with a sphere with the same properties of the
surrounding medium. Physically, we ascribe this fact to the
enhancement of the total cross section due to the magnetism in the
scatterer.

\section*{Acknowledgements}

The authors acknowledge the support of the Brazilian agencies CNPq
(303990/2007-4 and 476862/2007-8) and FAPESP (2008/02069-0).

\appendix
\section{Special functions}
\label{sec:special}

\subsection{Radial functions}
\label{sec:radial}

For the situation in which $m_i\not=0$ is verified, that is, the
imaginary part of relative refraction index (absorptive component)
is not null, we can write
    \begin{eqnarray}
        \int_0^ar^2\big|j_n(\rho_1)\big|^2{\rm
        d}r&=&\frac{a^3\big[y^*j_n(y)j_n'(y^*)-yj_n(y^*)j_n'(y)\big]}{y^2-y^{*2}}\nonumber\\
        &=&2a^3\big|j_n(y)\big|^2{\rm
        Re}\left[\frac{\varphi_n(y^*)}{y^2-y^{*2}}\right]\ ,\label{int-bes1}
    \end{eqnarray}
where $\rho_1=mkr$, $y=mka$, $\varphi_n(y)=y {\rm
d}_y[\ln\psi_n(y)]$. Eq.~(\ref{int-bes1}) is provided, in terms of
Bessel cylindrical functions, by Watson~\cite{watson}. In the
present context, to treat only with spherical Bessel functions, we
have used the relation
$({2\rho_1/\pi})^{1/2}j_n(\rho_1)=J_{n+1/2}(\rho_1)$
\cite{arfken}.

Also, if the relative refraction index $m$ is real, accordingly to
\cite{watson}, the integral in Eq.~(\ref{int-bes1}) can be simply
rewritten as
\begin{equation}
    \int_0^ar^2j_n^2(\rho_1){\rm
    d}r=\frac{a^3}{2}\big[j_n^2(y)-j_{n-1}(y)j_{n+1}(y)\big]\
    ,\label{int-bes2}
\end{equation}
which is obtained by taking the limiting $m_i\to 0$ in Eq.~(\ref{int-bes1}), and using L'Hospital rule and recurrence relations.

\subsection{Angular functions}
\label{sec:angular}

In the expansion of EM fields, it becomes natural to define the
angular functions $\pi_n(\cos\theta)
={P_n^1(\cos\theta)}/{\sin\theta}$ and $\tau_n(\cos\theta)={{\rm
d}P_n^1(\cos\theta)}/{{\rm d}\theta}$, where $\theta$ is the
scattering angle and $P_n^1$ is an associated Legendre function of
first order. These angular functions are quite convenient in the
calculation of fields intensities.

Due to properties involving the associated Legendre functions,
$\pi_n$ and $\tau_n$ satisfy the following expressions, $\forall
n,n' \in \N$:
\begin{eqnarray}
\int_{-1}^{1} {\rm d}(\cos \theta) \left(\pi_n   \pi_{n'} + \tau_n  \tau_{n'} \right) & = & \frac{2n^2(n+1)^2}{2n+1} \; \delta_{n,n'}\ ,  \label{ort1} \\
\int_{-1}^{1} {\rm d}(\cos \theta) \left(\pi_n \tau_{n'} + \tau_n    \pi_{n'} \right) & = & 0\ ,  \label{ort2} \\
\int_{-1}^{1} {\rm d}(\cos \theta)        \pi_n   \pi_{n'} \sin^2
\theta                  & = & \frac{2n(n+1)}{2n+1} \;
\delta_{n,n'} \label{ort3} \;.
\end{eqnarray}
These expressions facilitate the determination of quantities
involving fields intensities. We emphasize that, in the classical
books of scattering theory, Eq.~(\ref{ort3}) is not found in this
explicit form \cite{bohren,kerker,hulst}.

\section{Limiting cases}
\label{sec:limiting}

In these particular cases, we remark that $n=1$ is sufficient to
study the nonmagnetic scattering theory. Here, we imperatively
have to consider $n=1$ and $n=2$ to keep consistent orders in the
Mie coefficients.

\subsection{Small particle limit}

For the small argument limit into the Mie scattering coefficients,
we obtain
\begin{eqnarray}
    a_1&\approx&\frac{\imath x^3}{3} \; \frac{\varphi_1(mx)-2m\widetilde{m}}{\varphi_1(mx)+m\widetilde{m}} \nonumber \\
            &\ &-\frac{\imath x^5}{5} \; \frac{\left[\varphi_1(mx)-m\widetilde{m}\right]^2-m\widetilde{m}\varphi_1(mx)}{\left[\varphi_1(mx)+m\widetilde{m}\right]^2} \nonumber \\
            &\
            &+\frac{x^6}{9}\left[\frac{\varphi_1(mx)-2m\widetilde{m}}{\varphi_1(mx)+m\widetilde{m}}\right]^2+\mathcal{O}(x^7)\ ,\label{a-pe1}\\
    b_1&\approx&\frac{\imath x^3}{3} \; \frac{\varphi_1(mx)-2m/\widetilde{m}}{\varphi_1(mx)+m/\widetilde{m}}  \nonumber \\
            &\ &-\frac{\imath x^5}{5} \; \frac{\left[\varphi_1(mx)-m/\widetilde{m}\right]^2-(m/\widetilde{m})\varphi_1(mx)}{\left[\varphi_1(mx)+m/\widetilde{m}\right]^2} \nonumber \\
            &\
            &+\frac{x^6}{9}\left[\frac{\varphi_1(mx)-2m/\widetilde{m}}{\varphi_1(mx)+m/\widetilde{m}}\right]^2+\mathcal{O}(x^7)\ ,\\
    a_2&\approx&\frac{\imath x^5}{45} \; \frac{\varphi_2(mx)-3m\widetilde{m}}{\varphi_2(mx)+2m\widetilde{m}} + \mathcal{O}(x^7)\ ,\\
    b_2&\approx&\frac{\imath
    x^5}{45} \; \frac{\varphi_2(mx)-3m/\widetilde{m}}{\varphi_2(mx)+2m/\widetilde{m}} + \mathcal{O}(x^7) \; .
\end{eqnarray}

For the Mie internal coefficients, the approximations assume the
form below:
\begin{eqnarray}
    c_1&\approx&\frac{mx^2}{\psi_1(mx)}\frac{m/\widetilde{m}}{\left[\varphi_1(mx)+m/\widetilde{m}\right]}\\
        &\ &-\frac{mx^4}{\psi_1(mx)}\frac{(m/\widetilde{m})\left[\varphi_1(mx)-m/\widetilde{m}\right]}{2\left[\varphi_1(mx)+m/\widetilde{m}\right]^2}+\mathcal{O}(x^5)\ ,\\
    d_1&\approx&\frac{(m/\widetilde{m})x^2}{\psi_1(mx)}\frac{m\widetilde{m}}{\left[\varphi_1(mx)+m\widetilde{m}\right]}\\
        &\
        &-\frac{(m/\widetilde{m})x^4}{\psi_1(mx)}\frac{m\widetilde{m}\left[\varphi_1(mx)-m\widetilde{m}\right]}{2\left[\varphi_1(mx)+m\widetilde{m}\right]^2}+\mathcal{O}(x^5)\
        ,\\
    c_2&\approx&\frac{m}{3\psi_2(mx)}\frac{(m/\widetilde{m})x^3}{\left[\varphi_2(mx)+2m/\widetilde{m}\right]}+\mathcal{O}(x^5)\ ,\\
    d_2&\approx&\frac{(m/\widetilde{m})}{3\psi_2(mx)}\frac{m\widetilde{m}x^3}{\left[\varphi_2(mx)+2m\widetilde{m}\right]}+\mathcal{O}(x^5)\
    .\label{d-pe2}
\end{eqnarray}
Note that, for these approximations, the scattering coefficients
$a_1,\ a_2,\ b_1$ and $b_2$ have order
$\mathcal{O}\left[(ka)^{7}\right]$, whereas the internal
coefficients $c_1,\ c_2,\ d_1$ and $d_2$ are
$\mathcal{O}\left[(ka)^{5}\right]$. Terms for $n>2$ are ignored
here.

\subsection{Rayleigh approximation}

In this approximation, in which $|m|x\ll1$, the Mie scattering
coefficients can be write as
\begin{eqnarray}
    a_1&\approx& \frac{- 2 \imath x^3}{3} \; \frac{m\widetilde{m}-1}{m\widetilde{m}+2}  \nonumber \\
        &\ &-\frac{\imath
        x^5}{5} \; \frac{m^3\widetilde{m}-6m\widetilde{m}+(m\widetilde{m})^2+4}{(m\widetilde{m}+2)^2} \nonumber\\
        &\
        &+\frac{4x^6}{9}\left(\frac{m\widetilde{m}-1}{m\widetilde{m}+2}\right)^2+\mathcal{O}(x^7)\ ,\\
    b_1&\approx& \frac{- 2\imath
    x^3}{3} \; \frac{m/\widetilde{m}-1}{m/\widetilde{m}+2} \nonumber\\
        &\ &-\frac{\imath
        x^5}{5} \; \frac{m^3\widetilde{m}-6m/\widetilde{m}+(m/\widetilde{m})^2+4}{(m/\widetilde{m}+2)^2}  \nonumber\\
        &\
        &+\frac{4x^6}{9}\left(\frac{m/\widetilde{m}-1}{m/\widetilde{m}+2}\right)^2+\mathcal{O}(x^7)\ ,\\
    a_2&\approx& \frac{-\imath
    x^5}{15} \; \frac{m\widetilde{m}-1}{2m\widetilde{m}+3} + \mathcal{O}(x^7)\ ,\\
    b_2&\approx& \frac{-\imath
    x^5}{15} \; \frac{m/\widetilde{m}-1}{2m/\widetilde{m}+3} + \mathcal{O}(x^7)\
    ,
\end{eqnarray}
and the Mie internal coefficients assume the form
\begin{eqnarray}
    c_1&\approx&\frac{3}{2\widetilde{m}+m}\left[1+\frac{(mx)^2}{10}\right]\nonumber\\
        &\ &-\frac{3x^2}{2}\left[1+\frac{(mx)^2}{10}\right]\frac{(2\widetilde{m}-m)}{(2\widetilde{m}+m)^2}+\mathcal{O}(x^5)\ ,\\
    d_1&\approx&\frac{3}{2+m\widetilde{m}}\left[1+\frac{(mx)^2}{10}\right]\nonumber\\
        &\ &-\frac{3x^2}{2}\left[1+\frac{(mx)^2}{10}\right]\frac{(2-m\widetilde{m})}{(2+m\widetilde{m})^2}+\mathcal{O}(x^5)\ ,\\
    c_2&\approx&\frac{5}{m\widetilde{m}(3+2m/\widetilde{m})}+\mathcal{O}(x^5)\ ,\\
    d_2&\approx&\frac{5}{m(3+2m\widetilde{m})}+\mathcal{O}(x^5)\ .
\end{eqnarray}
Taking the particular case $m=\widetilde{m}$, Mie coefficients for
nonmagnetic scattering are recovered \cite{bohren}.

\subsection{Ferromagnetic limit for $x\ll1$}

This approximation, similar to Rayleigh limit, is derived directly
from the approximation of small spheres compared to wavelength.
Using the expressions for large argument limit present in
\cite{abramowitz}, one can obtain
\begin{eqnarray}
    a_1&\approx&\frac{\imath x^3}{3} \; \frac{x\tan(mx)+2\widetilde{m}}{x\tan(mx)-\widetilde{m}} \nonumber\\
            &\ &-\frac{\imath x^5}{5} \; \frac{\left[x\tan(mx)+\widetilde{m}\right]^2+\widetilde{m}x\tan(mx)}{\left[x\tan(mx)-\widetilde{m}\right]^2} \nonumber \\
            &\
            &+\frac{x^6}{9}\left[\frac{x\tan(mx)+2\widetilde{m}}{x\tan(mx)-\widetilde{m}}\right]^2+\mathcal{O}(x^7)\ ,\\
    b_1&\approx&\frac{\imath x^3}{3} \; \frac{\widetilde{m}x\tan(mx)+2}{\widetilde{m}x\tan(mx)-1} \nonumber \\
            &\ &-\frac{\imath x^5}{5} \; \frac{\left[\widetilde{m}x\tan(mx)+1\right]^2+\widetilde{m}x\tan(mx)}{\left[\widetilde{m}x\tan(mx)-1\right]^2} \nonumber\\
            &\
            &+\frac{x^6}{9}\left[\frac{\widetilde{m}x\tan(mx)+2}{\widetilde{m}x\tan(mx)-1}\right]^2+\mathcal{O}(x^7)\ ,\\
    a_2&\approx&\frac{\imath x^5}{45} \; \frac{x-3\widetilde{m}\tan(mx)}{x+2\widetilde{m}\tan(mx)} + \mathcal{O}(x^7)\ ,\\
    b_2&\approx&\frac{\imath
    x^5}{45}  \frac{\widetilde{m}x-3\tan(mx)}{\widetilde{m}x+2\tan(mx)} + \mathcal{O}(x^7) \;
    ;
\end{eqnarray}
the internal coefficients are
\begin{eqnarray}
    c_1 &\approx& \frac{ m}{  \cos(mx)}\frac{x^2}{\left[\widetilde{m} x\tan(mx) - 1   \right]}\nonumber\\
    &\ &-\frac{mx^4}{2\cos(mx)}\frac{\left[\widetilde{m}x\tan(mx)+1\right]}{\left[\widetilde{m}x\tan(mx)-1\right]^2}+ \mathcal{O}(x^5)\ ,\\
    d_1 &\approx& \frac{ m}{  \cos(mx)}\frac{x^2}{\left[ x\tan(mx) - \widetilde{m}   \right]}\nonumber\\
    &\ &-\frac{mx^4}{2\cos(mx)}\frac{\left[x\tan(mx)+\widetilde{m}\right]}{\left[x\tan(mx)-\widetilde{m}\right]^2} + \mathcal{O}(x^5)\ ,\\
    c_2 &\approx& \frac{-m}{3\cos(mx)}\frac{x^3}{\left[2              \tan(mx) + \widetilde{m}x\right]} + \mathcal{O}(x^5)\ ,\\
    d_2 &\approx& \frac{-m}{3\cos(mx)}\frac{x^3}{\left[2\widetilde{m} \tan(mx) +              x\right]} + \mathcal{O}(x^5)\; .
\end{eqnarray}

In this case, since low order in the size parameter is used, one
can obtain an analytical expression for the physical quantities,
such as cross sections, for instance.

\end{document}